\documentclass[letter]{aa}  
\usepackage{graphicx}
\usepackage{caption}
\usepackage{subcaption}
\usepackage[varg]{txfonts}
\usepackage{setspace}
\usepackage{booktabs}
\usepackage{multirow, makecell}
\usepackage{float}
\usepackage{mydblfloatfix} 

\usepackage{float}
\usepackage{tikz}
\usepackage{color}
\graphicspath{{./}{./figure/}}

\usepackage{xcolor}

\titlerunning{Magnetic helicity and energy budgets of jets from an
emerging active region}

\usetikzlibrary{arrows,positioning} 
\tikzset{
    >=stealth',
    punkt/.style={
           rectangle,
           rounded corners,
           draw=black, very thick,
           text width=6.5em,
           minimum height=2em,
           text centered},
    pil/.style={
           ->,
           thick,
           shorten <=2pt,
           shorten >=2pt,}
}
\newcommand\rectagular[1][red]{\begin{tikzpicture}
\draw [fill=red,red] (0.2,0.2) rectangle (0.3,0.3); 
\end{tikzpicture}
}\usepackage{lipsum}% http://ctan.org/pkg/lipsum

\begin{document}

\title{Magnetic helicity and energy budgets of jet events from an emerging
  solar active region}

    \author{A. Nindos\inst{\ref{inst1}} \and S. Patsourakos\inst{\ref{inst1}} \and K. Moraitis\inst{\ref{inst1}} \and V. Archontis\inst{\ref{inst1}} \and E. Liokati\inst{\ref{inst1}} \and M. K. Georgoulis\inst{\ref{inst2},\ref{inst3}} \and A. A. Norton\inst{\ref{inst4}} }

\institute{Section of Astrogeophysics, Department of Physics, University of Ioannina, 45110, Greece.\label{inst1} \\
\email{anindos@uoi.gr}
\and
The Johns Hopkins University Applied Physics Laboratory, Laurel, MD, 20723, USA \label{inst2}
\and
Research Center for Astronomy and Applied Mathematics, Academy of Athens, Athens, 11527, Greece\label{inst3}
\and
HEPL Solar Physics, Stanford University, 94305-4085, Stanford, CA, USA \label{inst4}
}
    
\date{Received date /
Accepted date }

\abstract
{
Using photospheric vector magnetograms obtained by the Helioseismic and Magnetic Imager onboard the Solar Dynamics Observatory and a magnetic connectivity-based method, we computed the magnetic helicity and free magnetic energy budgets of a simple bipolar solar active region (AR) during its magnetic flux-emergence phase, which lasted $\sim$47 hrs. The AR did not produce any coronal mass ejections (CMEs) or flares with an X-ray class above C1.0, but it was the site of 60 jet events during its flux-emergence phase. The helicity and free energy budgets of the AR were below established eruption-related thresholds throughout the interval we studied. However, in addition to their slowly varying evolution, each of the time profiles of the helicity and free energy budgets showed discrete localized peaks, with eight pairs of them occurring at times of jets emanating from the AR. These jets featured larger base areas and longer durations than the other jets of the AR. We estimated, for the first time, the helicity and free magnetic energy changes associated with these eight jets, which were in the ranges of $0.5-7.1 \times 10^{40}$ Mx$^2$ and $1.1-6.9 \times 10^{29}$ erg, respectively. Although these values are one to two orders of magnitude smaller than those usually associated with CMEs, the relevant percentage changes were significant and ranged from 13\% to 76\% for the normalized helicity and from 9\% to 57\% for the normalized free magnetic energy. Our study indicates that jets may occasionally have a significant imprint in the evolution of helicity and free magnetic energy budgets of emerging ARs.
} 
\maketitle                         

\section{Introduction} \label{sec: introduction}

Solar jets are collimated ejections of plasma that is launched outward
along magnetic field lines. They occur prolifically in diverse
environments such as coronal holes, the quiet Sun, and active regions
(ARs) and are observed in different wavebands. These are most notably soft X-rays,
EUV, and H$\alpha$; in the latter case, they are called surges \citep[see reviews
by][and references therein]{Raouafi2016,Shen2021,Schmieder2022}.

Jets are frequently associated with photospheric magnetic flux emergence and/or
cancellation, as well as with signatures of the impulsive release of energy such as
micro-flaring activity at their bases (in what follows, we use the term ``jet base'' to denote the structure in the EUV images from which the spire of the jet appears to emanate). These observations inspired a variety of
numerical models in which jets result from magnetic reconnection between the
emerged magnetic field and the preexisting open magnetic field lines
\citep[e.g.][]{Shibata1992,Yokoyama1995,Yokoyama1996,Archontis2008,Archontis2012,Archontis2013}. Sometimes jets appear in conjunction with the eruption of
mini filaments \citep[e.g.][]{Sterling2015,Sterling2016}, which motivated
\citet{Wyper2017,Wyper2018} to suggest that the material is ejected via a
breakout mechanism.

There is a tradition of studying energetic magnetic phenomena in terms of their
magnetic free energy (i.e., the term of the magnetic energy that is due to
electric currents) and helicity (i.e., a measure of the twist, writhe, and
linkage of the magnetic field lines) budgets. Older
\citep[e.g. see][ and references therein]{Pevtsov2014} and more recent
\citep[e.g.][]{Liokati2022,Liokati2023,Liu2023,Sun2024} results indicate
that ARs
tend to produce eruptive flares (i.e., flares accompanied with coronal mass
ejections; CMEs) when they accumulate significant budgets of both magnetic
free energy and helicity. Furthermore, some studies
\citep[e.g.][]{Pariat2017,Thalmann2019,Gupta2021} indicate that the ratio of the
magnetic helicity of the current-carrying field to the total magnetic
helicity (named helicity index by these authors) is a reliable
eruptivity proxy, whereas total magnetic energy and helicity are not.

Contrary to flares and CMEs, only a small number of publications
discuss how the occurrence of jets is related to the helicity budget of
their source regions. Numerical experiments \citep[see][]{Linan2018,Pariat2023}
indicate
that jet-producing simulations contain higher values of both free
magnetic energy and helicity than the ones with no eruption and no jet. More
importantly, it was found that the jet may occur when the helicity index
attains its maximum value. As far as observations are concerned,
\citet{Green2022} monitored the evolution of the helicity index in a large
emerging eruptive active region and found that some major jets occurred at
time intervals when the helicity index obtained large values.

\begin{figure}[!ht]
\centering
\includegraphics[width=9cm]{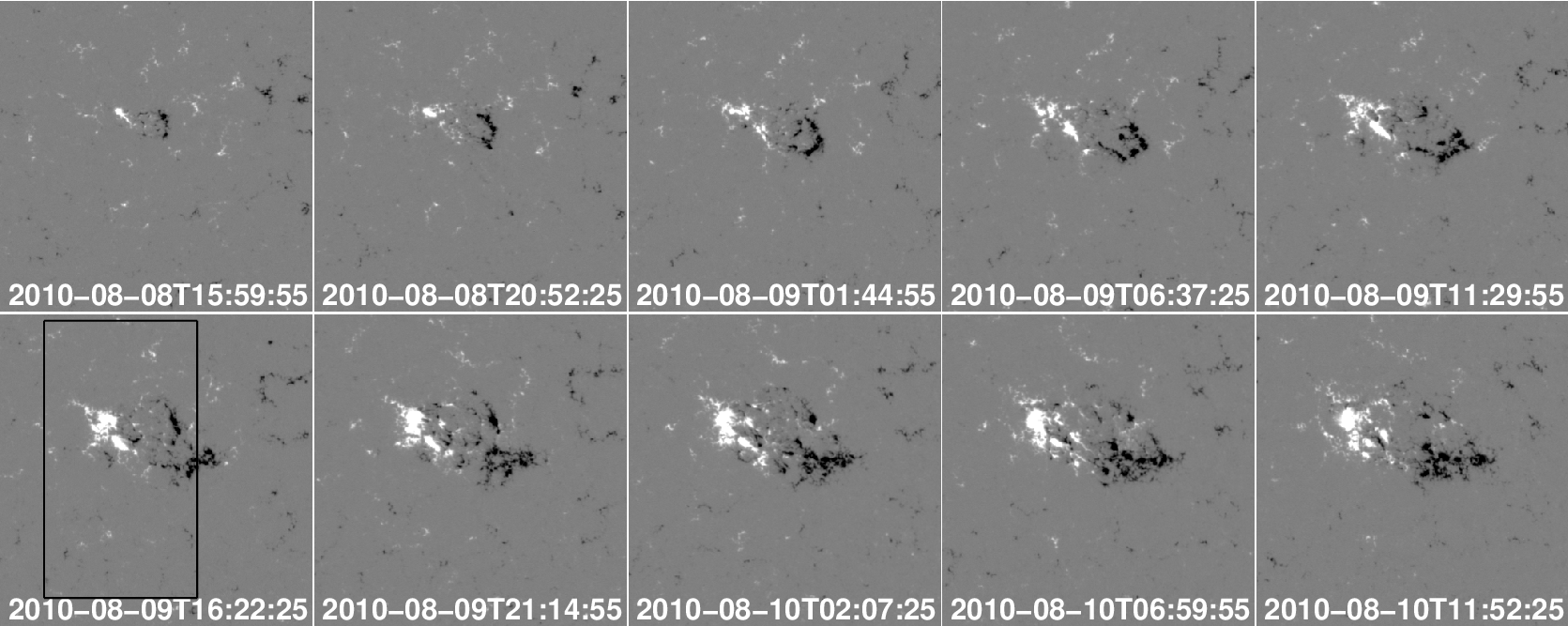}
\caption{Selected HMI line-of-sight magnetograms showing evolution of photospheric magnetic field in AR11096. In each image the field of view is $205\arcsec \times 205\arcsec$. The black box shows the area that is displayed in Fig. A.1.}
\end{figure}

In this letter, we present the evolution of magnetic helicity and free magnetic
energy in an AR that produced several jet events during its flux-emergence phase. We show, for the first time with such clarity, that several
of these jets occurred at times when both magnetic helicity and free energy
showed distinct localized peaks, and we calculate the helicity and free magnetic
energy changes associated with these jets. 

\section{Observations and data reduction} 

In this letter, we study the emerging AR NOAA 11096. The morphological evolution
of its photospheric magnetic field was studied using line-of-sight magnetograms
from the Helioseismic and Magnetic Imager (HMI) onboard
the Solar Dynamics Observatory (SDO). The magnetic helicity and free magnetic
budgets of the AR were calculated using HMI vector magnetograms. More
specifically, we used the ``hmi.sharp\_cea\_720s\_dconS'' data series, which
provides Lambert cylindrical-equal-area (CEA) projections of the photospheric
magnetic field vector \citep[][]{Bobra2014} that were corrected for scattered
light \citep[][]{Couvidat2016,Norton2018}. The correction did not modify the morphology of
the AR field captured by the magnetograms, but it increased the average total
field strength across the field of view by a factor that varied during the
observations from 1.19 to 1.41. The pixel size of the magnetograms was
equivalent to about 360 km at the disk center, while the cadence of the vector field
image cubes was 12 min.

At each timestamp we computed the instantaneous free magnetic energy and
helicity budgets using the connectivity-based (CB) method of
\citet{Georgoulis2012}. The input of this method is a single vector magnetogram that is
partitioned to yield a connectivity matrix populated by the magnetic flux
associated with connections between partitions of opposite polarities. This
collection of connections is treated as an ensemble of force-free flux tubes;
each with known footpoints, force-free parameter, and flux. For the system of
these flux tubes, the method delivers lower-limit estimates of their free
magnetic energy and helicity.

The results from the CB method were compared with results from the
flux-integration (FI) method; in it, the magnetic helicity and energy fluxes
across the photospheric boundary are computed. The inputs of this method are
the normal and tangential components of the photospheric magnetic field and the cross-field velocity field at the photosphere
\citep[e.g.][]{Kusano2002}. Details of the computational procedure are given
in \citet{Liokati2022}.

The jets associated with the AR were identified in 211 \AA\ images
from the Atmospheric Imaging Assembly (AIA) telescope onboard SDO. This AIA
channel is sensitive to 2 MK plasmas. Since it is unlikely that tiny,
short-lived jets have any impact on the magnetic helicity and energy budgets
of the AR, we degraded the cadence of our AIA datacube from 11 s to 2
min.

\section{Results}

The emergence of AR11096 started on 08 August, 2010 11:00 UT (heliographic
coordinates N22W08) in an area without preexisting ARs (see
Fig. 1 for characteristic snapshots). Inspections of 211-\AA\ AIA movies
indicate
that the AR produced several jets, most of which occurred during its emergence
phase. Therefore, we limited our calculations to the flux-emergence phase of the
AR, which (as was found from the time profile of the unsigned flux of the
photospheric field) lasted $\sim$47 hours.

Figure 1 indicates that AR11096 was a simple bipolar AR. The AR produced neither
CMEs nor flares with an X-ray class above C1.0 (see Liokati et al. 2022 and
the time profile of the 211-\AA\ flux from the AR in Fig. 2(a)).
During the interval we studied, several jets emanated from the AR,
most of them from its eastern part (see Appendix A). We identified 60 jets,
which is
probably a lower limit because the cadence of the AIA data we used was 2 min.
Characteristic snapshots of major jets hosted by the AR are given in Fig. A.1.

\begin{figure}[!hb]
\centering
\includegraphics[width=9cm]{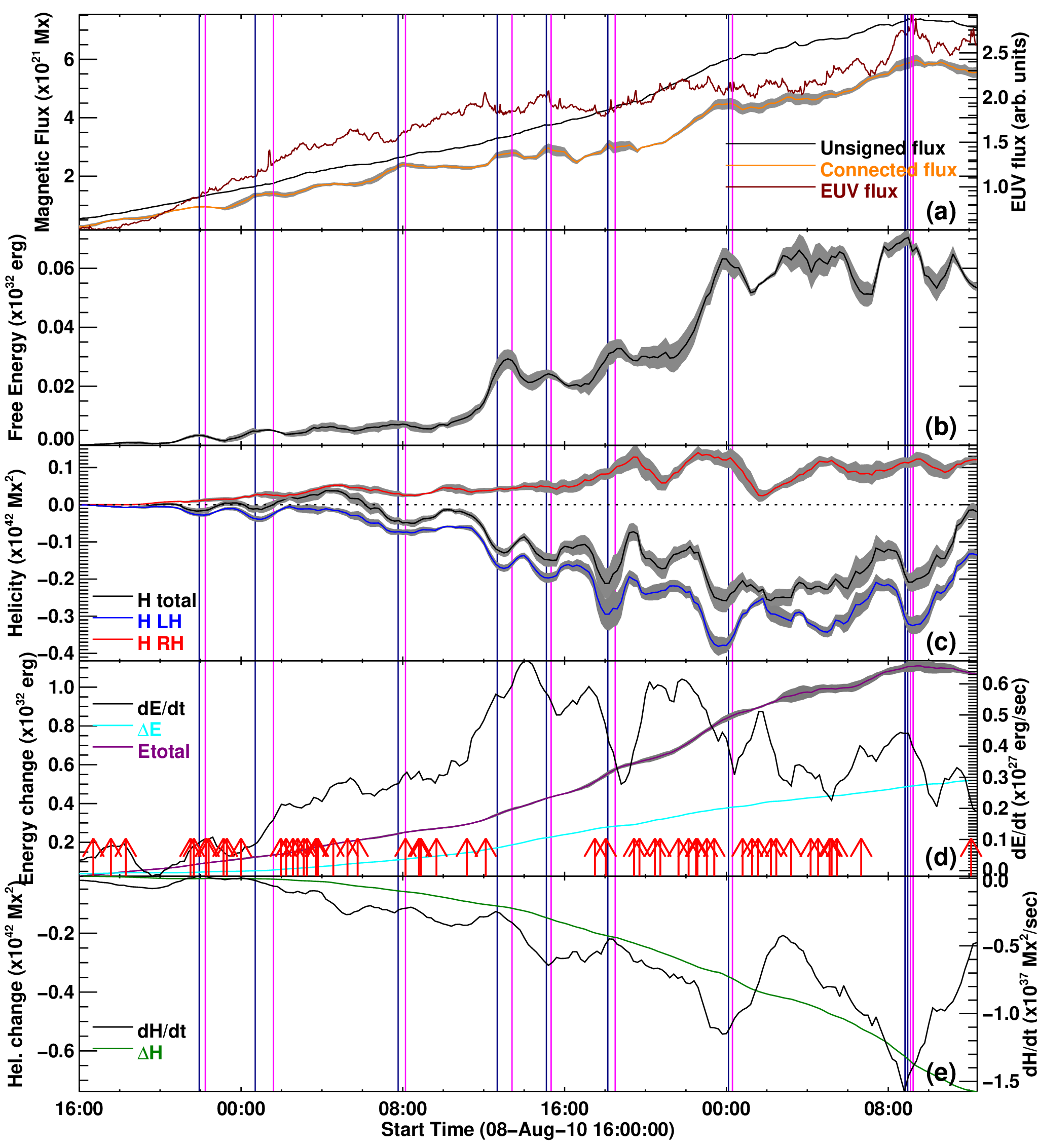}
\caption{Evolution of magnetic properties of AR11096. (a) Unsigned magnetic flux, unsigned connected magnetic flux used in the CB method, and EUV flux of the AR resulting from AIA's 211 \AA\ channel (black, yellow, and maroon curves, respectively). (b) Free magnetic energy. (c) Net, left-handed, and right-handed helicity (black, blue, and red curves, respectively). (d) Total magnetic energy from the CB method, magnetic energy injection rate from the FI method, and the corresponding accumulated magnetic energy (purple, black, and cyan curves, respectively). (e) Helicity injection rate from the FI method and the corresponding accumulated helicity (black and green curves, respectively). Vertical dark blue and pink lines show the start and end times of major jet events, while arrows indicate the start time of the remaining jets of the AR. Error bars are indicated by the gray bands.}
\end{figure}

\begin{figure}[!hb]
\centering
\includegraphics[width=9cm]{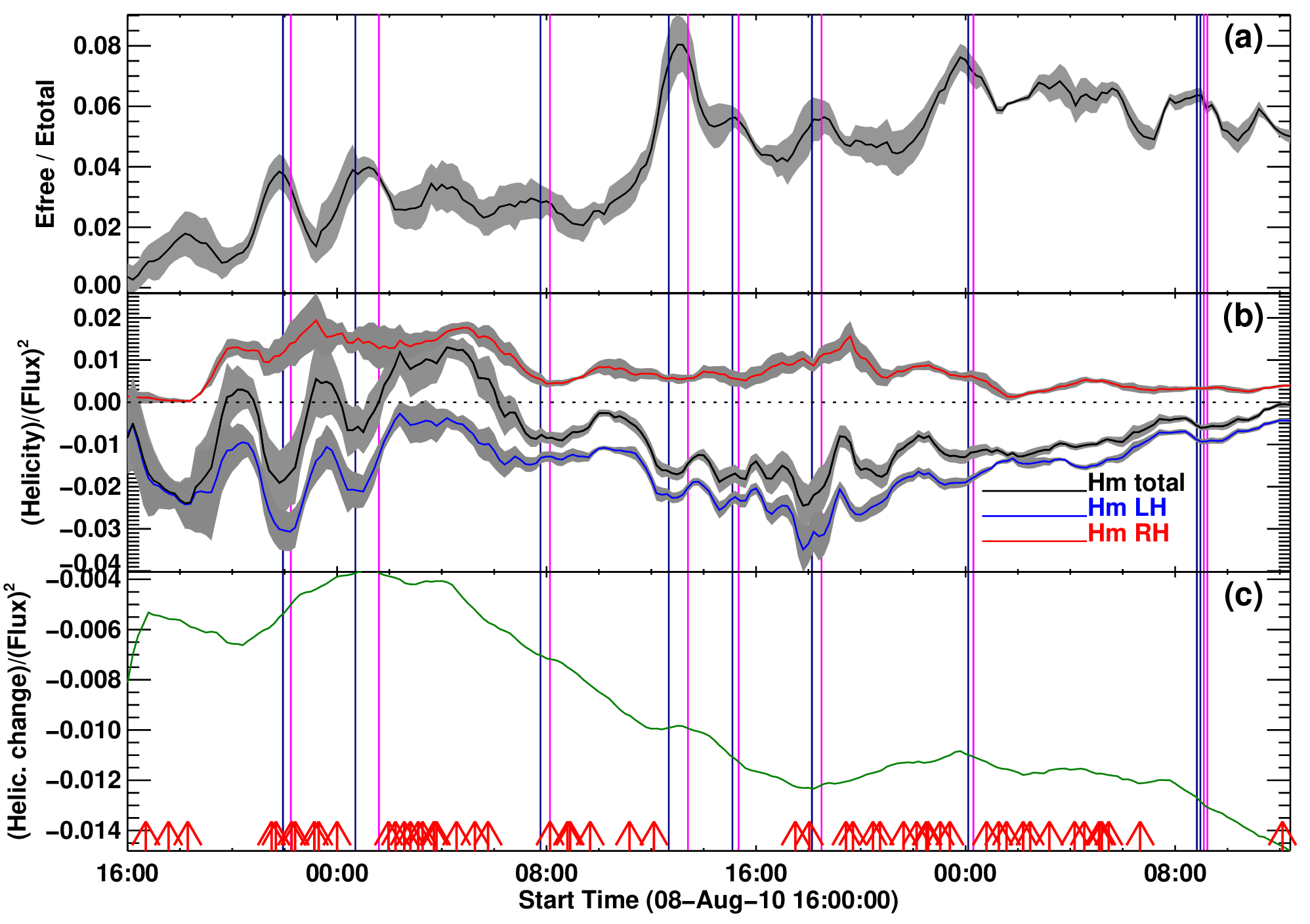}
\caption{Evolution of normalized magnetic quantities for AR11096. (a) Ratio of the free magnetic energy to the total magnetic energy. (b) Ratios of the net, left-handed, and right-handed helicity to the connected magnetic flux squared (black, blue, and red curves, respectively). (c) Ratio of the accumulated helicity from the FI method to the connected magnetic flux squared. Vertical lines
and arrows are as in Fig. 2.}
\end{figure}

In Fig. 2 we give the time profiles of the free magnetic energy ($E_f$, panel
b) and magnetic helicity (panel c) of the AR. In order to evaluate the long-term
evolution of the magnetic energy and helicity, all pertinent curves in this
letter are 48 min averages of the actual curves. Panels (b) and (c) indicate
that both the free magnetic energy and net helicity curves exhibit slowly
varying evolutionary patterns that are consistent with the evolution of both
the unsigned magnetic flux, $\Phi$, of the AR and the so-called connected flux,
$\Phi_{\mathrm{conn}}$
\citep[i.e., the magnetic flux that populates the connectivity matrix employed by the CB method; see][]{Georgoulis2012}.

The helicity curves (panel c) show that for most of the time interval we studied
the net helicity is negative, in agreement with the hemispheric helicity rule
\citep{Pevtsov1995}; there is only an $\sim$5 hour intrusion of positive net
helicity starting at $\sim$09 August 01:40 UT. Therefore, it is not a surprise
that the net helicity curve by and large follows the evolution of the negative
helicity curve (correlation coefficient of 0.92).

The evolution of the injection rates of magnetic energy, $dE/dt$, and helicity,
$dH/dt$, from the FI method and the resulting accumulated quantities
($\Delta E$ and
$\Delta H$, respectively) are given in panels (d) and (e) of Fig. 2. A direct
comparison between the results from the CB-method and the results from the
FI-method is not possible
\citep[see the discussion in][]{Thalmann2021,Liokati2023}.
However, Fig. 2 indicates that there is a very good
resemblance between the evolution of $\Delta E$ and the evolution of the
total magnetic energy, $E_{tot}$, derived from the CB method (correlation
coefficient of 0.96). The correlation coefficient between $\Delta H$
and the net helicity from the CB-method is 0.75 (i.e., rather strong but weaker
than that of the $\Delta E$-$E_{tot}$ pair). Furthermore, comparisons of the
maximum values of the pertinent budgets derived by the two methods reveal
differences of factors of 2.2 to 2.9 (these comparisons are meaningful because
the starting value of both methods is close to zero). We repeated the calculations of the $E_f$ and $H$ budgets by the two methods after we degraded the resolution of the magnetograms to 2$\arcsec$. The new curves largely preserve the jet-related local peaks, while the differences between the two methods decrease by factors of 1.1-1.3. This behavior is consistent with the findings of \citet{Wang2022}. However, we note that the interpolation of the vector field data may influence the solenoidality of the field and the quality of azimuth disambiguation.

The free magnetic energy and helicity budgets of the AR are always clearly
below reported thresholds for the occurrence of major flares (compare the values
appearing in panels (b) and (c) of Fig. 2 with the thresholds of $4 \times
10^{31}$ erg and $2 \times 10^{42}$ Mx$^2$, respectively, established by
\citet{Tziotziou2012}). The accumulated budgets of $\Delta E$ and $\Delta H$
resulting from the FI-method (see panels (d) and (e) of Fig. 2) are also lower
than the corresponding thresholds established by \citet{Liokati2022} ($2
\times 10^{32}$ erg and $9 \times 10^{41}$ Mx$^2$, respectively), despite the
fact that the data used by \citet{Liokati2022} were not corrected for
scattered light.

The slowly varying trends of the free magnetic energy and helicity budgets are
paired with shorter localized peaks (see Fig. 2 (b, c)). In eight cases these
peaks are synchronized with jets produced in the AR. This is evident in Fig. 2,
where the start and end times of major jet events produced in the
AR are marked by dark blue and pink vertical lines, respectively.
The localized peaks associated with the occurrence of jets appear co-temporally
in the free magnetic energy, net
helicity, $H$, and left-handed helicity ($H_{LH}$) curves. With the probable
exception of the
seventh event, their signature is not prominent in the right-handed helicity
($H_{RH}$)
curve (that is, the minority sense of helicity). Furthermore, all the
$E_f$-$H$ localized peaks occurring in conjunction with jets stand out beyond
error bars \citep[the latter are indicated by the gray bands of Fig. 2 and
result from the standard deviations of the moving five-point averages of the
pertinent curves; see][]{Moraitis2021,Liokati2023}.

In Fig. 2 there are nine pairs of colored vertical lines corresponding to eight
localized peaks (hereafter referred to as events 1-8) in each of the $E_f$, $H$,
and $H_{LH}$ curves. The mismatch comes from the fact that event 8 takes place
during the occurrence of two temporally overlapping jets (see
bottom right panel of Fig. A.1). Most of these
localized $E_f$ and $H$ peaks occur either around the start time of a major jet
(peaks 1, 3, and 5) or between the start and end times of a major jet (events 2 and
4), while for the others, small temporal offsets can be registered between the
$E_f-H$ peaks and the interval of occurrence of the jet. These offsets were on the order
of 12-24 min, and consequently they were barely resolved because the cadence of
the magnetograms was 12 min.

Panels (d) and (e) of Fig. 2 indicate that the occurrence of events 1-8
was not associated with any prominent signature in the $\Delta E$ and
$\Delta H$ curves. However, the
$dE/dt$ curve shows local peaks associated with events 3, 4, and 8 while the
$dH/dt$ curve shows absolute-value local peaks associated with events 3, 5, 7,
and 8. Using the binomial distribution test we found that the likelihood
for incidental peak matchings between the curves from the two methods is
11\% and 24\% for the magnetic energy and helicity, respectively.

In Fig. 3 we present the evolution of the ratio of $E_f$ to the total magnetic
energy, $E_f/E_{tot}$, as well as the connected-magnetic-flux normalized
helicities ($H/\Phi_{\mathrm{conn}}^2$, $H_{RH}/\Phi_{\mathrm{conn}}^2$, $H_{LH}/\Phi_{\mathrm{conn}}^2$).
The values of these curves are a factor of $\sim$2 lower than those associated
with the two large eruptive ARs studied by \citet{Liokati2023}.
%However, events 1-8 are all associated with well-defined localized peaks of the free-energy and helicity normalized parameters.
However, the normalized parameters of the free energy and helicity
  exhibit well-defined local peaks that are associated with jet events
  1-8. We cannot test the behavior of the
helicity index (see Sect. 1) because the CB method does not allow the
decomposition of the total helicity to the parameters required for its
calculation (the helicity index is calculated on 3D-inferred helicity
estimation, hence the CB method is not designed to compute it).

Returning to the CB budgets appearing in Fig. 2, the free magnetic
energy and helicity changes ($DE_f$ and $DH$, respectively)
associated with events
1-8 were calculated as the difference between the relevant localized peak and
the value at the curve's point of inflection occurring just after the
localized peak.
The results appear in columns two and four of Table 1. The corresponding
percentages of the normalized $E_f$ and $H$ losses (see Fig. 3) are given in
columns three and five of Table 1. These percentages are all negative,
implying that free energy and helicity are taken away by the jets. The same
behavior has been registered for large flares \citep[e.g.][]{Wang2023,Liokati2023}.
In this respect, jets 1-8 can be considered as miniature eruptions.

In addition to the jet events that were associated with localized
peaks in the free magnetic energy and helicity budgets of the AR, several other
jets were also produced in the AR (their start times are denoted by the arrows
in Figs. 2 and 3).
The duration of these jets was short (see Fig. A.2(b)), and neither of
them coincided with localized peaks appearing in both the $E_f$ and $H$ time
profiles. We investigated how the former group of jets could be further
distinguished from the latter. To this end, we calculated the apparent
area of the bases (hereafter referred to as ``area of the bases'')
of all jets that emanated from AR11096 (see Appendix A for details). The results
appear in Fig. A.2(a), which indicates that, on average, the areas of
the bases of
the jets that are co-temporal with local peaks in the $E_f$-$H$ budgets of the
AR are statistically larger than the areas of those that have no
significant imprint in the evolution of the $E_f$-$H$ budgets.

\section{Conclusions}

In this work we used the CB method \citep[][]{Georgoulis2012} to evaluate the
free magnetic energy and helicity budgets of a small emerging AR. The AR was
bipolar and during its emergence phase (which lasted 47 hours) produced no flares
above C1.0-class or CMEs. However, we were able to identify 60 jets emanating from the AR by visually
inspecting 211-\AA\ movies. 

Throughout the interval we studied, the $E_f$ and $H$ budgets of the AR were
below established thresholds \citep[see][]{Tziotziou2012,Liokati2022};
if these are crossed, the AR is likely to erupt. The time profiles of the free
magnetic energy and helicity resulted from the superposition of two components:
(i) a slowly varying one that was broadly consistent with the evolution of
both the total unsigned magnetic flux of the AR and the connected magnetic
flux; and (ii) discrete localized peaks of much shorter durations (full widths at
half maximum of $\sim$55-160 min). Eight such peaks in each of the $E_f$
and $H$ curves (all well beyond uncertainties)
occurred co-temporally with jet events produced in the AR. These
local helicity peaks can reasonably be attributed to the jet events because no
other type of eruptive activity was registered in the AR. Furthermore, their
pairing with $E_f$ localized peaks supports the same conclusion for the origin
of the simultaneous $E_f$ peaks.

The jets associated with localized peaks in the $E_f$ and $H$
budgets of the AR are distinguished from the other AR jets by the larger
areas of their bases and their longer durations. The former
is in line with the fact that $E_f$ and $H$
are extensive quantities. It is interesting, though, that these major
jets are also associated with local peaks in the time profiles of the
normalized free magnetic energy and helicity parameters.

The free magnetic energy and helicity losses associated with the jets
are in the ranges of
$(1.1-6.9) \times 10^{29}$ erg and $(0.5-7.1) \times 10^{40}$ Mx$^2$,
respectively. These values are one to two orders of magnitude smaller
  than the relevant changes associated with CMEs \citep[see][and references
    therein]{Liokati2023}.
The derived free energy losses are consistent with the high-end
limits of the thermal energy of jets \citep[see][]{Shen2021}. There are no
previous explicit reports based on observations about helicity changes
directly associated with
jets. The percentage losses associated with the jets are significant: 9-57\% for
the normalized free magnetic energy and 13-76\% for the normalized helicity.
There is a trend for the percentage losses to be larger early on in the evolution
of the AR (see Table 1).

This is the first report where changes in the magnetic free energy and helicity
budgets of an AR are registered with such clarity with jet activity.
\citet{Green2022} was the first to report jet activity in intervals when the
helicity index (see Sect. 1) attains large values. In our study, the close
synchronization of localized $E_f$ and $H$ peaks with jet events allowed us to
estimate, for the first time, the free magnetic energy and helicity budgets
associated with individual jet events. 

More case studies are required to check how often jets may have significant
imprints in the evolution of the free magnetic energy and helicity budgets of
emerging ARs. As more case studies accumulate, it will be interesting to
investigate whether the collective $E_f$ and $H$ budgets of such jets are
significant. Computations of the field line helicity \citep[e.g.][]{Yeates2018,Moraitis2019,Moraitis2021,Moraitis2024} may also provide insights into the distribution of helicity over the different components of individual jets.

\begin{table}
\begin{center}
\caption{Free magnetic energy and helicity budgets of jet events.}
\begin{tabular}{ccccc}
\hline
Event  & \multicolumn{2}{c}{$D E_f$}  & \multicolumn{2}{c}{$D H$} \\
Number & ($\times 10^{29}$ erg) & (\%) & ($\times 10^{40}$ Mx$^2$) & (\%) \\
\hline
1      &   -1.9       &  -57    &  1.3 & -76 \\
2      &   -1.4       &  -26    & 0.5 & -40 \\
3      &   -1.1       &  -15    & 1.7 & -35 \\
4      &   -7.9       &  -27    & 3.9 & -31 \\
5      &   -3.9       &  -15    & 4.0 & -27 \\
6      &   -2.9       &  -9    & 5.6 & -26 \\
7      &   -5.5       &  -18    & 3.4 & -13 \\ 
8      &   -6.9       &  -14    & 7.1 & -34 \\
\hline
\end{tabular}
\end{center}
%}
\end{table}

\begin{acknowledgements}
We thank the referee for their constructive comments.
AN, SP, KM, VA, and AAN acknowledge support by the ERC Synergy Grant (GAN:
810218) ``The Whole Sun''. AN thanks Robert H. Cameron and Allan Sacha Brun for
useful discussions.
\end{acknowledgements}

\bibliographystyle{aa-note} %% aa.bst but adding links and notes to references
\bibliography{ms_v2}

\begin{appendix}

\section{Jets produced in AR11096}

\begin{figure}[!h]
\centering
\includegraphics[width=9cm]{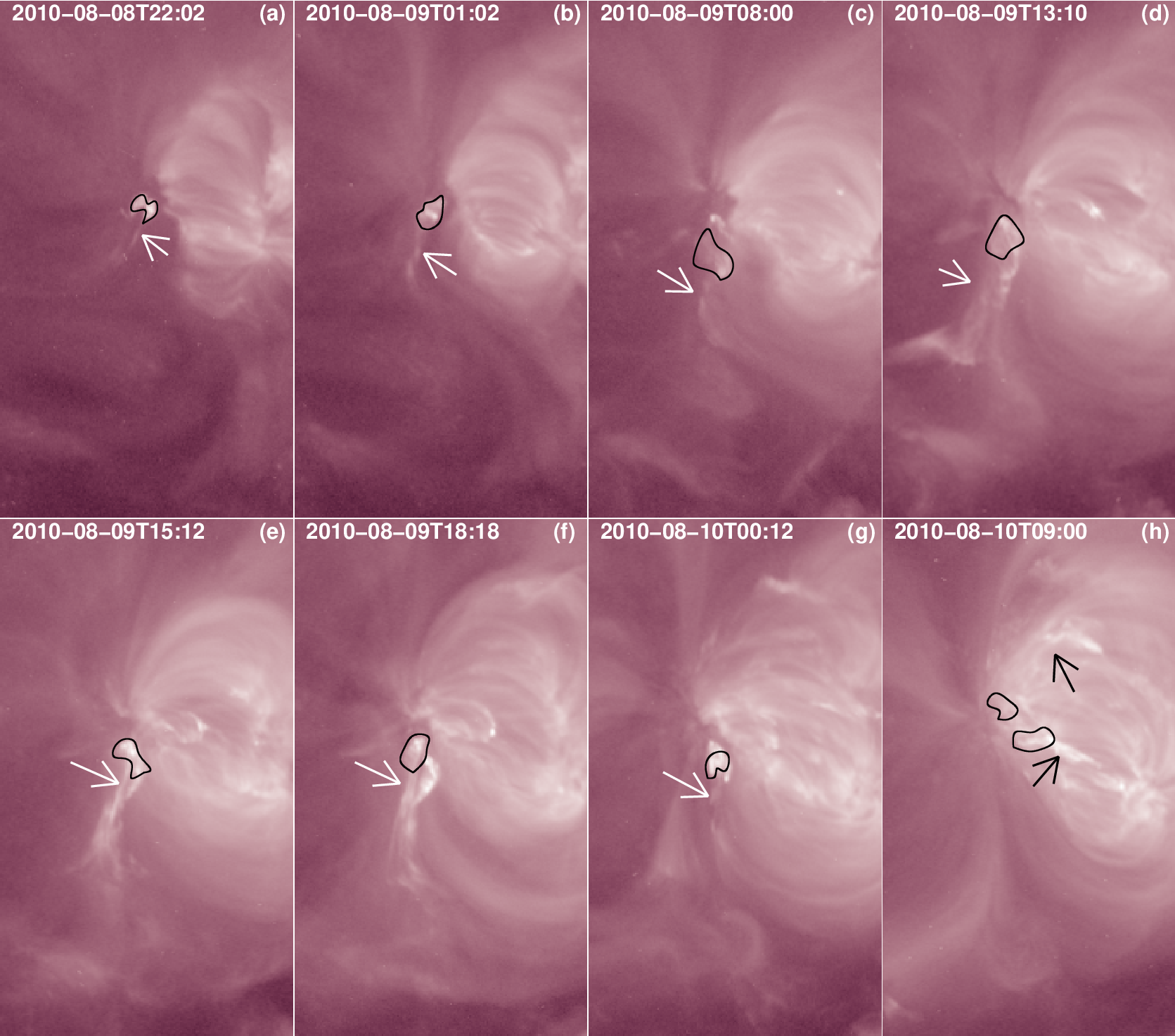}
\caption{Selected 211-\AA\ AIA images showing the jets (denoted by
  arrows) that were associated with localized peaks of the free magnetic
  energy and helicity budgets of AR11096. Contours outline the
   calculated apparent areas of the bases of the jets (see text for details).
  The field of view is $103\arcsec \times
  181\arcsec$ and corresponds to the black box of Fig. 1. The associated movie
  is available online.}
\end{figure}

In Fig. A.1 we show characteristic snapshots of the major jets that
were
associated with localized peaks of the free magnetic energy and helicity
budgets of
AR11096. These jet events are marked with arrows.
All of the jets that occurred in the AR are marked with arrows in the
movie that accompanies the paper.
Note that there
are two arrows in the bottom right panel of Fig. A.1 due to the presence of two
temporally overlapping
jet events. All of the jets presented in Fig. A.1 occurred in the eastern part of the AR.
In more detail, jets 1 and 2 (see panels a and b) emanated from approximately
the same location and the same is true for jets 3-7 (see panels c-g).

The calculations of the apparent areas of the bases of all jets that we detected
in AR11096 are presented in panel (a) of Fig. A.2. In each snapshot
the jet base area was computed by
employing a $35\arcsec \times 35\arcsec$ box just below the spire of the jet
and by taking into account only the box pixels with intensities exceeding the
$2\sigma$ levels of their
time series above the background (for the snapshots of Fig. A.1 the resulting areas are outlined by the black contours).
In addition to this semi-automatic procedure,
the bases of the jets were determined by visual inspection. Both methods
yielded consistent results (differences of $\sim$30-40\%). The areas that appear
in Fig. A.2 result from the average values derived from the two methods.
In Fig. A.2 the red symbols correspond to the events displayed in Fig. A.1
while the black symbols corresond to the other jets. Each vertical error bar
in the figure denotes the standard deviation of the computed time series of
the base area of the jet. When only one image of a jet was available
(that was the case for the majority of events) no error bar was attached to
our calculation.

In panel (b) of Fig. A.2 we show the duration of the jets. The duration is defined as the interval between the last and first appearance of the jet spire in the AIA images. The duration of jets appearing in only one AIA image has been somewhat arbitrarily set to 2 min (i.e., the cadence of our AIA datacube).

\begin{figure}[h]
\centering
\includegraphics[width=9cm]{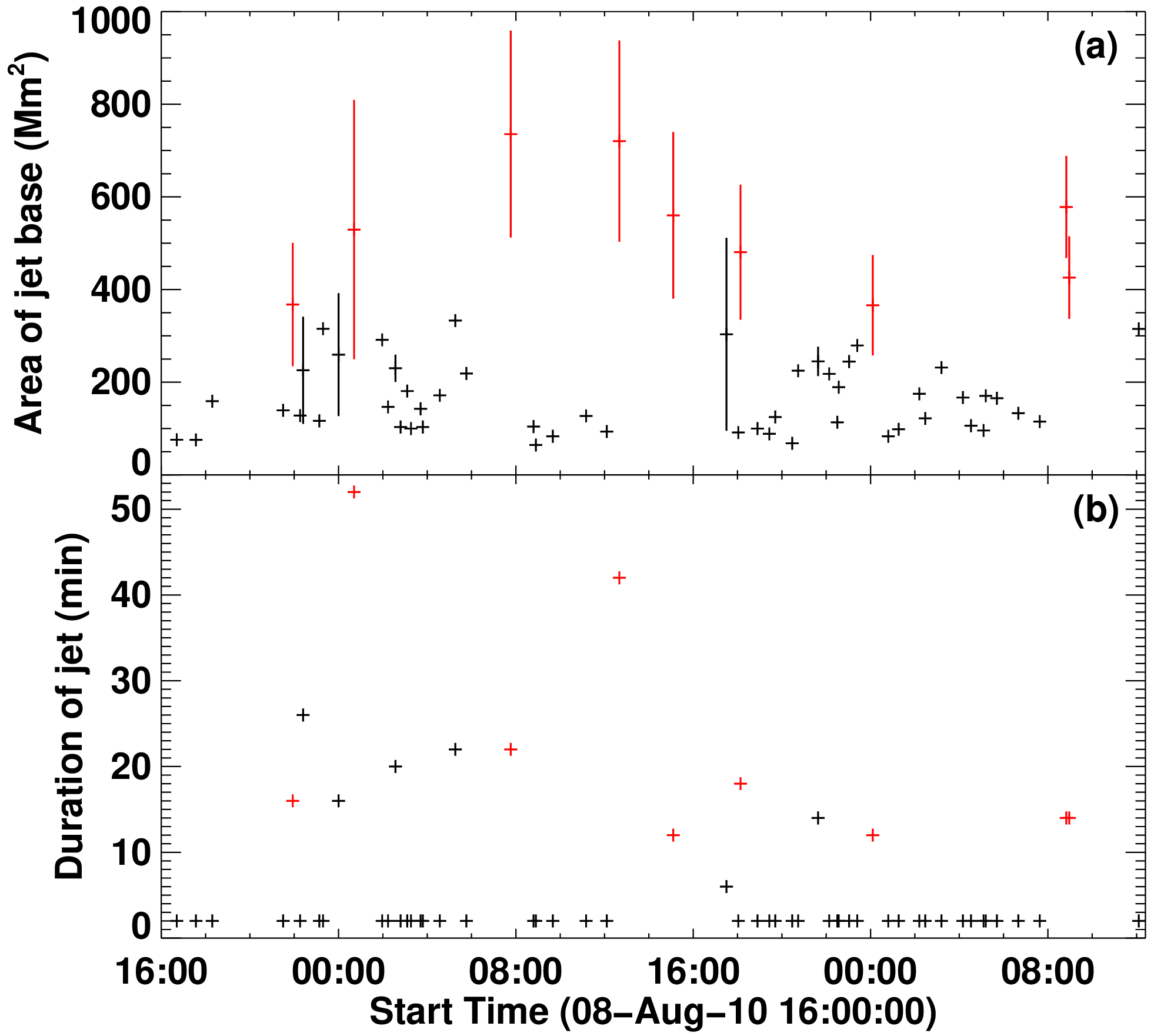}
\caption{(a) Apparent areas of the bases of all jet events that we detected in
AR11096. When possible the measurements are
accompanied with their uncertainties (see text for more details).
(b) Duration of jets. The duration of those jets that appeared in only one AIA image has been set to 2 min. In both panels red symbols
correspond to the events presented in Fig. A.1 while black symbols are used for
the other events. In both panels each symbol has been placed at the
start time of the corresponding jet.}
\end{figure}

\end{appendix}

\end{document}